\documentclass[12pt]{article}
\usepackage{amsmath}
\usepackage{graphicx,psfrag,epsf}
\usepackage{enumerate}
\usepackage{amsfonts}
\usepackage[utf8]{inputenc}
\usepackage[T2A]{fontenc}

\newcommand{\blind}{0}

\newtheorem{definition}{Definition}

\usepackage[backend=bibtex,
style=authoryear,
natbib=true,
maxcitenames=2,
giveninits=true,
maxbibnames=99]
{biblatex}
\begin{document}

\def\spacingset#1{\renewcommand{\baselinestretch}%
{#1}\small\normalsize} \spacingset{1}


\if0\blind
{
  \title{\bf funLOCI: a local clustering algorithm for functional data}
  \author{Jacopo Di Iorio \hspace{.2cm}\\
    Penn State University, Dept. of Statistics, PA, USA\\
    jqd5830@psu.edu
    \\
    \\
    Simone Vantini \\
    MOX - Politecnico di Milano \\
    simone.vantini@polimi.it}
  \maketitle
} \fi

\if1\blind
{
  \bigskip
  \bigskip
  \bigskip
  \begin{center}
    {\LARGE\bf funLOCI: a local clustering algorithm for functional data}
\end{center}
  \medskip
} \fi

\bigskip

\begin{abstract}
Nowadays, more and more problems are dealing with data with one infinite continuous dimension: functional data. In this paper, we introduce the funLOCI algorithm which allows to identify functional local clusters or functional loci, i.e., subsets/groups of functions exhibiting similar behaviour across the same continuous subset of the domain. The definition of functional local clusters leverages ideas from multivariate and functional clustering and biclustering and it is based on an additive model which takes into account the shape of the curves. funLOCI is a three-step algorithm based on divisive hierarchical clustering. The use of dendrograms allows to visualize and to guide the searching procedure and the cutting thresholds selection. To deal with the large quantity of local clusters, an extra step is implemented to reduce the number of results to the minimum.

\end{abstract}

\section{Introduction}
\label{intro}
One of the fundamental need in data mining is to group a given set of objects according to some measure of similarity or dissimilarity. Trying to accomplish this mission, the scientific community has created a wide number of algorithms and methods. The most famous of them are known under the name of clustering methods, usually applicable to data arranged in a data matrix.
The main element of clustering is the similarity between rows or columns of the data matrix. This procedure leads to the discovery of some similarity groups at the expense of obscuring other similarity groups: it is not possible to have row groups and column groups at the same time. Indeed, considering the two dimensions of data matrices, rows and columns, observations and features, one can obtain groups of similar observations according to all the features or, instead, groups of similar features according to all the observations. 
Furthermore, most of these algorithms seek a disjoint cover of the set of elements, i.e. they require that two cluster groups can not overlap and that each element, row or column, must be clustered into exactly one group. To overcome these limitations, a large number of algorithms that perform overlapping clustering (fuzzy clustering) and simultaneous clustering on both the dimensions of the data matrix has been proposed. 

In recent years, thanks to the possibility of collecting and storing increasingly larger amount of information, the statistical community started to be interested in solving problems whose data were characterized by a very large dimension $p$. A special and extreme case of this type of data is functional data (i.e. data that can be represented as curves, surfaces, $\dots$). Functional datasets are modelled as samples of random variables which take values in an infinite-dimensional functional space, e.g., a space of functions defined on some set $T$, for instance, time interval.
Functional data clustering received particular attention from statisticians in the last decade. In their survey, Jacques and Preda (\citeyear{jacques2014functional}) categorize all the different clustering approaches into four groups: raw-data methods, filtering methods, adaptive methods, and distance-based methods.

The raw-data methods comprise all the techniques coming from the multivariate world. These methods cluster discretized versions of functional data using multivariate clustering methods without reconstructing the functional form. These early approaches are unable to take into account typical functional features such as continuity and derivatives. 

In the filtering methods, the high dimensionality of data is tamed by a filtering step which approximates the curves by means of a finite basis of expansions. After this first step, usually performed with B-splines or FPCA (see \cite{ramsay2007applied} for a general framework), multivariate clustering algorithms are used to define clusters of functional data. 
For instance, some algorithms propose k-means clustering on b-splines coefficients or principal component scores, while others apply unsupervised neural networks to Gaussian coefficient's basis.

The adaptive methods collect contributes that consider the basis expansion as a random variable having a cluster-specific probability distribution, instead of simple parameters. For this reason, most of these methods are based on probabilistic modelling of basis expansion or of some FPCA scores.

Finally, distance-based methods try to adapt popular geometric clustering algorithms, such as k-means and hierarchical clustering, to the functional setting. For these techniques, it is necessary to define new specific distances or dissimilarities between functions. Depending on the definition and computation of these measures, the methods belonging to the last group can be also related to either raw-data or filtering methods.

All these techniques consider the curves globally, over their entire domain, missing grouping that may occur only on a portion of it. To consider this scenario, one should refer to the recent sparse functional clustering literature or ``functional feature selection'', which it is the functional counterpart of the multivariate "domain selection'' problem, usually referred to as ``feature selection''. These methods are capable of clustering the data while also selecting their most relevant features for classification. In Floriello et al. (\citeyear{floriello2017sparse}), for instance, the functional sparse clustering is analytically defined as a variational problem with a hard thresholding constraint which ensures the sparsity of the solution. More recently, curve alignment was integrated in the sparse clustering procedure (\cite{vitelli2019novel}).  The ability to focus on subsets of the domain is a common characteristic with our work.

Precisely, in this paper, we define functional local clustering which allows to identify functional local clusters or functional loci, i.e., subsets/groups of functions exhibiting similar behaviour across the same continuous subset of the domain. The definition is based on an additive model which takes into account the shape of the curves. We also propose the new functional local clustering identifier (funLOCI) algorithm, a three step algorithm that permits the identification of functional local clusters.

This article is structured as follows: the definition and the validation of a functional local cluster is introduced in Section \ref{definitions}; in Section \ref{funLOCI} the funLOCI algorithm is presented; finally, to explain the practical usefulness of this new functional data analysis method, two case studies, one with simulated data and the other with real data, are shown in Section \ref{casestudies}.

\section{Definition of Functional Local Cluster}
\label{definitions}
Before identifying functional local clusters, we need to define what a functional local cluster is. A functional local cluster or functional locum is as a subset of functions, or curves, that exhibit similar behaviour (similar shape) across the same continuous subset of the domain $T$. Consistently with the existing literature, the concept of similarity has a central role.

Taking inspiration from the most comprehensive formulation of model-based clustering and multivariate biclustering, it is possible to give the following definition of functional local cluster:

\begin{definition}
A functional local cluster $Q = (I,S)$ is a subset $I$ of functions paired with a sub-interval $S$ of the domain $T$ s.t.
\begin{equation}
\label{functionalbic}
    f_{i}(t) = \mu^Q + \alpha_{i}^Q + \beta^Q(t) + \varepsilon_{i}(t) \hspace{10px} \forall f_{i} \in I \hspace{5px}and\hspace{5px} t \in S,
\end{equation}
where $f_{i}(t)$ is a general curve belonging to the local cluster $Q$, $\mu^Q$ is the specific mean of the local cluster $Q$, $\alpha_{i}^Q$ is the function-specific adjustment and $\beta^{Q}(t)$ is the $t$-varying pattern of the local cluster $Q$, and $\varepsilon_{i}(t)$ is an error term.
\end{definition}

To define uniquely the model parameters and make them identifiable, it is customary to impose the following constraints:
$\sum_{i \in I}\alpha_{i}^{Q} = 0$ and $\int_{S}\beta^{Q}(t)dt = 0$.  

Starting from Equation \ref{functionalbic}, one can define simpler kinds of loci such as the constant values local clusters and the constant values on rows/columns ones. 
\begin{figure}
    \centering
    \includegraphics[width=0.6\textwidth]{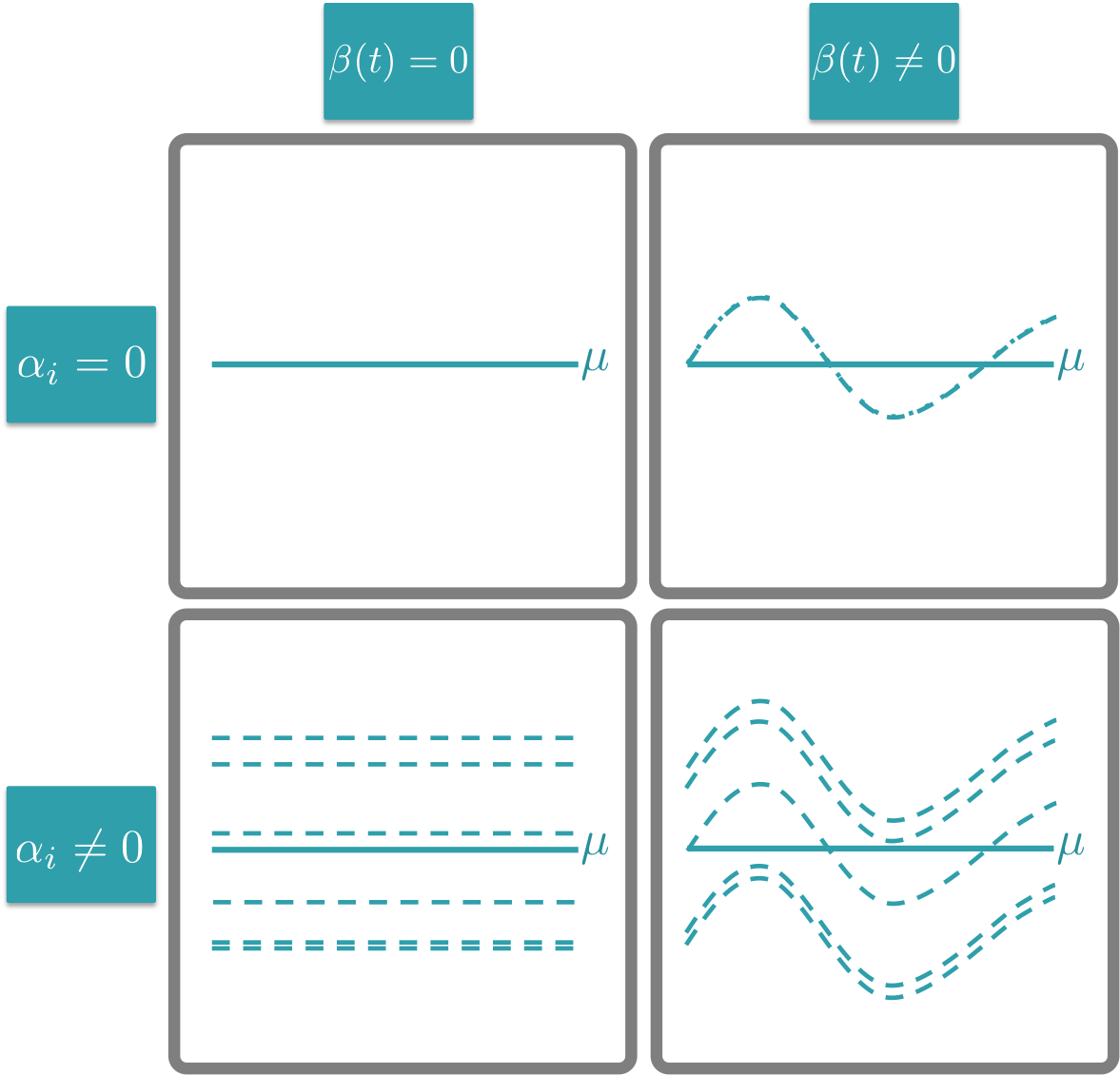}
    \caption{Different settings of $\alpha_i$ and $\beta(t)$ define different kinds of functional local clusters.}
    \label{fig:fourtypes}
\end{figure}
For instance, setting $\alpha_{i}^{Q} = 0$ and $\beta^{Q}(t) = 0$, the aforementioned expression is simplified in $f_{i}(t) = \mu$. The resulting locum is a constant local cluster and it is composed only by functions constantly equal to a given value $\mu^Q$ on a sub-interval $S$ of the time domain (top-left panel of Figure \ref{fig:fourtypes} ).

Setting $\alpha_{i}^{Q} \ne 0$ and $\beta^{Q}(t) = 0$, the obtained bicluster, expressed by $f_{i}(t) = \mu + \alpha_{i}$, is the constant values on rows bicluster. It is composed by parallel constant functions on $S$, a sub-interval of $T$ (bottom-left panel of Figure \ref{fig:fourtypes}).
On the other hand, setting $\beta^{Q}(t) = 0$ instead of $\alpha_{i}^{Q}$, the found local cluster is constant on columns and it is based on $f_{i}(t) = \mu^{Q} + \beta^{Q}(t)$. It consists of identical functions (top-right panel of Figure \ref{fig:fourtypes}).

Therefore, the complete formulation expressed in Equation \ref{functionalbic} identifies group of parallel non-necessarily constant functions on a sub-interval of the domain sharing the same shape (bottom-right panel of Figure \ref{fig:fourtypes}).


\subsection{Perfect Local Cluster}
From Equation \ref{functionalbic}, we define a perfect ideal local cluster as a local cluster having the error term $\varepsilon_{i}(t) = 0$. We can measure $\varepsilon_{i}(t)$, and so the adherence to the additive model (which it is unknown a priori), by using a validation score. We opted for a functional version of the mean squared residue score or H-score, introduced by Cheng and Church in their biclustering seminal paper (\citeyear{chengchurch}). A similar functional version can be found in the funCC functional biclustering algorithm proposed by Torti et al. (\citeyear{torti2022general}).

From Equation \ref{functionalbic}, the mean squared residue score for a functional local cluster $Q = (I,S)$ can be written in the following way:

\begin{equation}
\label{hscore}
    H(Q) = H(I,S) = \frac{1}{\mid I \mid}\frac{1}{\mid S \mid}\sum_{i \in I}\int_{S}(f_{i}(t) - (\hat{\mu}^Q + \hat{\alpha}_{i}^Q + \hat{\beta}^Q(t)))^{2}dt.
\end{equation}

In details, $f_{i}(t)$ is the value of function $i$ at instant $t$, $\mid I \mid$ is the cardinality of the set $I$, and $\mid S \mid$ is the length of the interval $S$. $\hat{\mu}^Q$, $\hat{\alpha}_{i}^Q$ and $\hat{\beta}^{Q}(t)$ are the estimates of $\mu^Q$, $\alpha_{i}^Q$, and $\beta^{Q}(t)$, and they are respectively defined as:

\begin{equation}
        \hat{\mu}^Q = f_{IS} = \frac{1}{\mid I \mid}\frac{1}{\mid S \mid}\sum_{i \in I}\int_{S}f_{i}(t)dt
\end{equation}

\begin{equation}
        \hat{\alpha}_{i}^Q = f_{iS} - \hat{\mu}^Q = \frac{1}{\mid S \mid}\int_{S}f_{i}(t)dt \hspace{5px} - \hat{\mu}^Q
\end{equation}

\begin{equation}
        \hat{\beta}^{Q}(t) = f_{I(t)} - \hat{\mu}^Q = \frac{1}{\mid I \mid}\sum_{i \in I} f_{i}(t) \hspace{5px} - \hat{\mu}^Q
\end{equation}

where $f_{iS}$ is the integral mean of the function $i$ in the sub-interval $S$, $f_{I(t)}$ is the sample mean of all the functions in $I$ at the time instant $t$ and $f_{IS}$ is the general mean of all the curves in $I$ in the whole sub-interval $S$ (sample mean of the integral means).

Considering the relationships between $\hat{\mu}^Q$, $\hat{\alpha}^{Q}_{i}$, $\hat{\beta}^{Q}(t)$, and $f_{I(t)}$, $f_{IS}$, $f_{iS}$, it is possible to write Equation \ref{hscore} as follows:

\begin{equation}
\label{hscore_practioners}
    H(Q) = H(I,S) = \frac{1}{\mid I \mid}\frac{1}{\mid S \mid}\sum_{i \in I}\int_{S}(f_{i}(t) - f_{iS} - f_{I(t)} + f_{IS})^{2}dt.
\end{equation}

The mean squared residue score for functional local cluster is then a measure of coherence used to validate a candidate locum by computing its adherence to the additive model. The optimum is given by the lowest score $H(I,S)=0$, a situation that is visually represented by perfect parallel curves. Modifying the model expressed in Equation \ref{functionalbic} obliges to also modify the mean squared residue score used in the detection. Setting $\alpha=0$, $\beta=0$ or both, the optimum is still represented by the lowest score $H(I,S)=0$.

\section{Looking for a functional local cluster}
\label{funLOCI}
\begin{figure}
\centering
\includegraphics[width=0.75\textwidth]{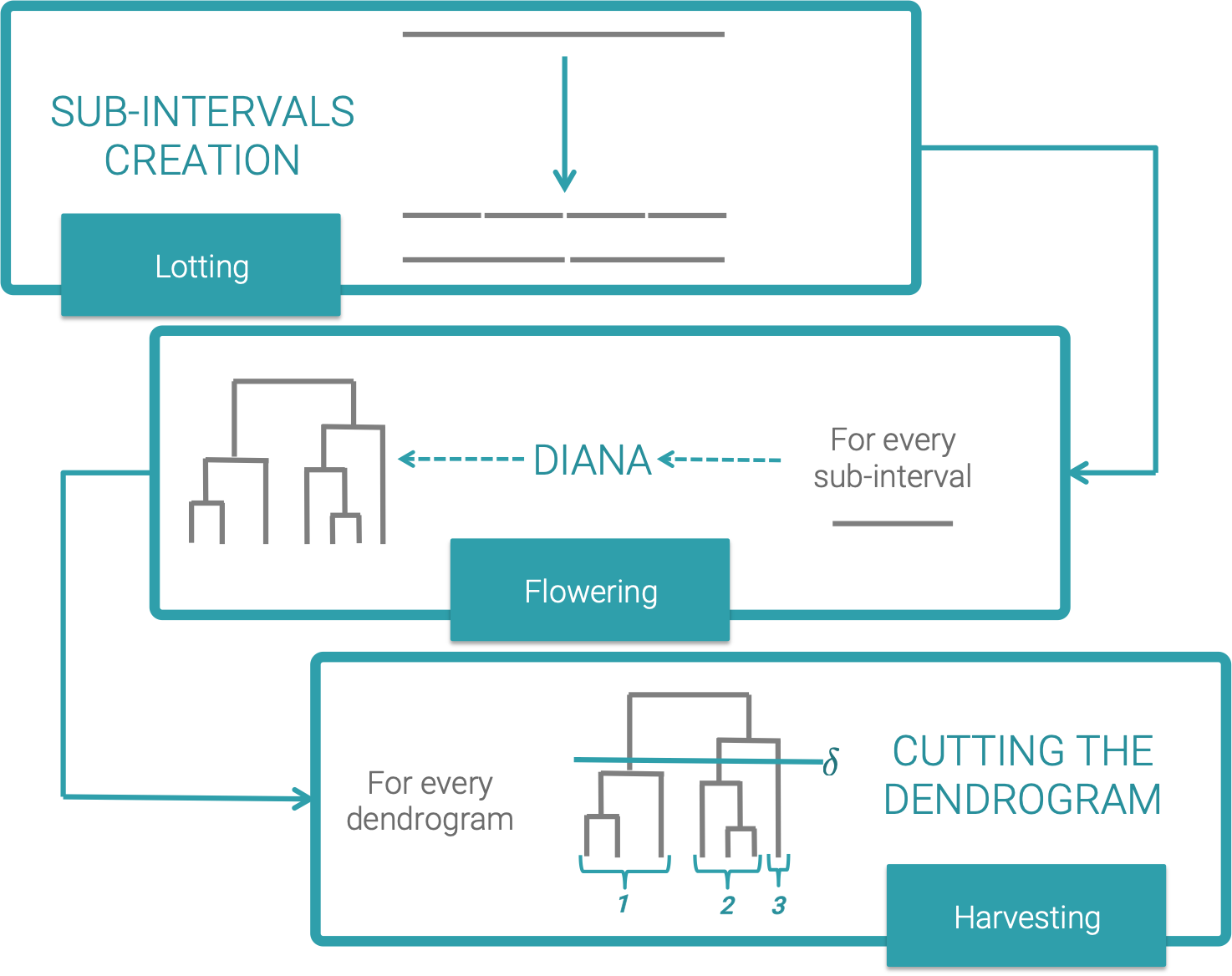}
\caption{funLOCI can be summarized in three steps performed one after the other: \textit{Lotting}, \textit{Flowering}, and \textit{Harvesting}. The first step aims at creating sub-intervals $S_{w}$ of the whole domain $T$. In the second step, called \textit{Flowering}, a hierarchical clustering algorithm based on the mean squared residue score is applied. In the \textit{Harvesting} step, the last one, all the candidate local clusters are collected. We propose the following routine.}
 \label{fig:algotable}
\end{figure}

Using the mean squared residue score it is possible to validate the goodness of a candidate locum. However, the additive models defining local clusters are unknown a priori, so we need an algorithm to find local clusters within the data. Many clustering and biclustering algorithms face the same issue using a brute-force approach. This approach is NP-hard in the worst case, but easily parallelizable. 
Similarly, funLOCI follows a brute-force strategy when dealing with the continuous dimension. Precisely, it is a three-step iterative procedure (Figure \ref{fig:algotable}) where each step is demanded to a specific task.

The first step, named \textit{Lotting}, aims at creating sub-intervals $S_{w}$ of the whole domain $T$. In the second step, called \textit{Flowering}, a hierarchical clustering algorithm based on the mean squared residue score is applied. It returns for every set of sub-curves (i.e., curves defined in the same sub-interval $S_{w}$) a hierarchy of candidate local clusters. In the \textit{Harvesting} step, the last one, all the candidate local clusters are collected. We propose the following routine.

\paragraph{Step 1 - Lotting}
Starting from a discrete grid based on the continuous dimension, the user decides the minimum length $c$ of the continuous interval to analyze. All the sub-intervals $S_{w} :\hspace{1px} \mid S_{w} \mid \geq c$ are considered.

Due to the computational complexity of this strategy, the \textit{Lotting} procedure also gives to the user the possibility to consider only a selection of all the sub-intervals.
The user can define $\bar{a} = \{a_1, \dots, a_n \}$, a vector of sub-intervals lower boundary positions, and $\bar{c} = \{c_1, \dots, c_n \}$, a vector with sub-intervals lengths. Then, every sub-interval $S_{w}$ is created in the following way: $S_{w} = ( \hspace{1px} a_{i}, min(a_{i} + c_{j}, \mid T\mid) \hspace{1px})$ for $\forall i,j$ where $\mid T \mid$ is the length of the domain $T$.

\paragraph{Step 2 - Flowering}
Focusing on the sub-functions defined on $S_{w}$, sub-interval of the domain $T$, a mean squared score based DIANA (DIvisive ANAlysis Clustering by \cite{kaufman2009finding}) algorithm is performed.
The DIANA algorithm is the most famous divisive hierarchical clustering algorithm. Initially, all data are in one cluster; an iterative procedure is then used to split the largest cluster into two parts. The splitting procedure is performed in the following way: DIANA chooses the cluster with the largest diameter, i.e., the maximum average dissimilarity; it initiates a new group called ``splinter group'' with the observation having the largest average dissimilarity from the other ones of the selected cluster; it reassigns to the splinter group all observations that are more similar to the new cluster than to the original one. The result is a division of the selected cluster into two new clusters. The algorithm is over when each cluster contains only a single observation

In our case the dissimilarity matrix used by DIANA is based on the $H_{score}$.
For each couple of curves $i,j \in I$, the dissimilarity $d_{ij}$ between $i$ and $j$ is $H(\big\{i,j\big\},S_{w})$, i.e., the $H_{score}$ of $(\big\{i,j\big\},S_{w})$, that is the matrix composed only by the two curves $i$ and $j$ evaluated in $S_{w}$: 

\begin{equation}
\label{d_i}
    d_{ij} = H(\big\{i,j\big\},S_{w}).
\end{equation}

Therefore, $d_{ij}=0$ means that $H(\big\{i,j\big\},S_{w})=0$, i.e., the curves $i$ and $j$ together compose a perfect local cluster in accordance to the additive model.

The result of this step is presented in a dendrogram that summarizes the DIANA top-down splitting procedure. Because the dissimilarity matrix is based on the mean squared residue score, we use the H-score as the height/vertical axis of the dendrogram.

The procedure aforementioned is performed for every sub-interval $S_{w}$. 
 
\paragraph{Step 3 - Harvesting} All the candidate local clusters are collected by cutting each dendrogram. The collection can be performed according to two different cutting strategies.

In the first one, the user decides a threshold for the H-score called $\delta$-threshold that it is going to be used in every sub-inteval $S_{w}$. The value $\delta$ is then used to cut the dendrograms and, consequently, to identify $\delta$-local clusters $Q$, i.e., loci having $H(Q)\leq\delta$. However, the $\delta$-threshold could be too severe in some sub-intervals, resulting in a large number of local clusters composed of just few sub-functions, and too succinct in some other, generating one unique functional local cluster with all the curves. 

In the second strategy the user select a $\delta$-percentage, $\delta^{\%}$, $0<\delta^{\%}<1$. Then the dendrogram defined in every $S_{w}$ is cut using a $\delta$-threshold $\delta_w = \delta^{\%} H(X,S_{w})$, where $H(X,S_{w})$ is the H-score of the local cluster composed of the totality of sub-curves $X$ defined in the sub-interval $S{w}$. Therefore, the resulting local cluster $Q$ has $H_{Q} \leq \delta^{\%} H(X,S_{w})$. In this way, the proposed threshold is more tailored to $S_w$. 

In order to select the most proper $\delta$ or $\delta^{\%}$ we can use a method similar to the elbow method. So, the best $\delta^{\%}$ parameter, for instance, is identified as the point where diminishing returns are no longer worth the additional cost. In our case, it corresponds to a point where using another $\delta^{\%}$ does not longer worth the additional cost in terms of number of local clusters, average number of elements per clusters, or average H-score. However, this ``elbow'' cannot always be unambiguously identified. This strategy can be applied to select a unique threshold for all the sub-intervals or to select a proper threshold for each sub-interval leading to more flexible results by applying the elbow method to each sub-interval.

\bigbreak
The three steps aforementioned, the \textit{Lotting}, the \textit{Flowering} and the \textit{Harvesting} are extremely customizable. This makes the whole algorithm very flexible and easily modifiable according to the problem under exam.

After the \textit{Harvesting} step, a set of candidate local clusters is identified. Even if the best strategy would be to check them one by one with the advice of a problem domain expert, when the minimum length $c$ is particularly small and all the possible sub-intervals are considered, the number of results could be overwhelming. Therefore, it could be useful to provide a way to order the results by importance and to reduce the number of local clusters: this is performed by the optional step of the \textit{Tasting}.

\paragraph{Step 4 - Tasting} In accordance with the multivariate biclustering literature, the most interesting results are the ones having the lowest H-score and the maximum dimension (typically in terms of the number of rows and columns) here in terms of sub-interval length $|S_{w}|$ and number of curves $|I|$. Although, in some applications one could be interested in small and punctual phenomena considering low dimensions. 
Due to the brute-force approach, our algorithm can find loci that are nested into some others or that are shifted overlapping versions of other ones. 

The local cluster $(K, S_{w_1})$ is nested inside the local cluster $(Y, S_{w_2})$  if $K \subseteq Y$ and $S_{w_1} \subseteq S_{w_2}$. Then, the local cluster $(K,S_{w_1})$ can be removed since it is totally included in $(Y, S_{w_2})$ and hence not informative. 

Instead, we define, the local cluster $(K, S_{w_1})$ as a shifted overlapping version of the local cluster $(Y, S_{w_2})$ if $K \subseteq Y$,  $\mid S_{w_1} \mid \leq \mid S_{w_2} \mid$, and $\mid S_{w_1 \cap w_2} \mid \geq 0.5 \mid S_{w1} \mid$, where $S_{w1 \cap w2} = S_{w1} \cap S_{w2}$ is the intersection between the two intervals. Then, the locum $(K, S_{w_1})$ is partially (more than 50$\%$ of its sub-interval length) inside $(Y, S_{w_2})$ and hence not completely informative and then removable. We want to remark that according to this definition nested loci are a particular case of shifted overlapping ones (with $\mid S_{w_1 \cap w_2} \mid = \mid S_{w_1} \mid$).

\begin{figure}
\centering
\includegraphics[width=0.8\textwidth]{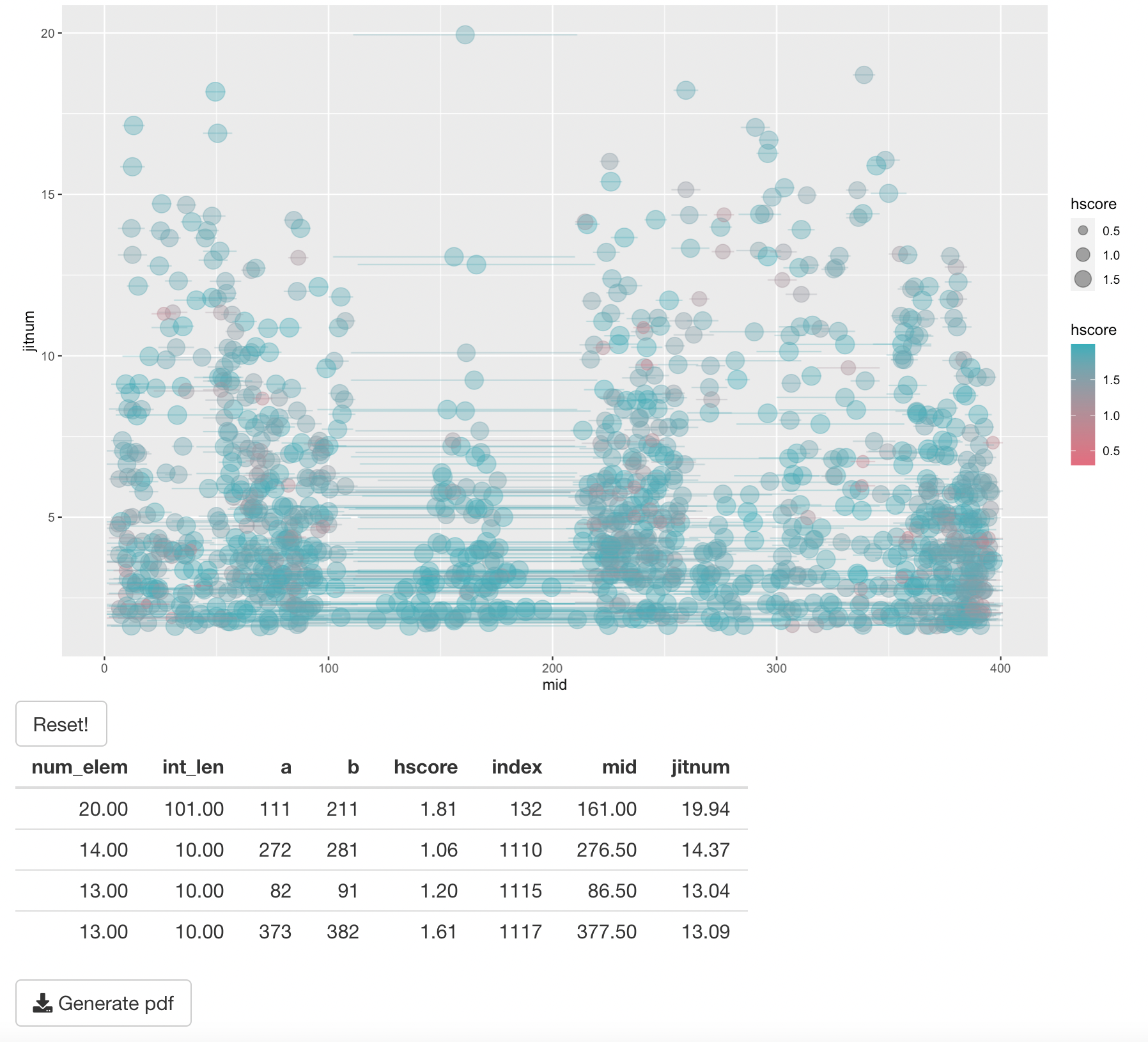}
\caption{The interactive plot works as a tool to help the user to explore manually all the results. Every candidate functional local cluster is represented by a dot crossed by a segment. The dot is the mid-point of the local cluster sub-interval while its color and its size are based on the local cluster H-score. The user can explore and filter the results by zooming and clicking directly on the dots. The table below the scatterplot shows the details about the clicked points. Clicking on the``Generate pdf'' button a pdf with the line plots of every selected locum is downloaded.}
 \label{fig:tastingexplorer}
\end{figure}

To get rid of nested and shifted overlapping local clusters and just highlight the most interesting ones, the \textit{Tasting} uses the following approach: the candidate local clusters are ordered according to sub-interval length $|S_{w}|$ and number of curves $|I|$ in decreasing order, and then H-score (increasingly). Then, the first locum - the most interesting one - is going to be the one with the longest sub-interval having the highest number of curves, and the lowest H-score. We set the first locum as ``interesting''.
All the ordered loci are processed sequentially and compared with the previous interesting ones. If the locum under analysis is nested inside another interesting locum or if it is a shifted overlapping version of another one, then it is considered ``not interesting''.
In this way, the \textit{Tasting} simplifies the revision of the results by reducing the number of local clusters to minimal. 
Even in this way, the results might be too many to allow the direct visual inspection of every locum one by one. For this reason, we suggest to drive the inspection by ordering the results according to the feature or features that are more important in the analysis. Visualization and interactive visualization can also play a pivotal role in the exploration of the local clusters. For this reason we implemented an interactive plot that can help in the manual identification of the best loci according to one's demand (see Figure \ref{fig:tastingexplorer}). All the loci are represented as dots crossed by a segment. The dot is the mid-point of the sub-interval while the color and the size of the dot are based on the H-score of the local cluster. The segment, instead, represents the sub-interval. The user can explore and filter the results by zooming and clicking directly on the dots. We are currently working on adding new features to the interactive visualization tool that can ease the manual and visual exploration.

\section{Case Studies}
\label{casestudies}
In this section, two different case studies are presented, in order to highlight the practical usefulness of the funLOCI algorithm. The first case is based on simulated data while the other one on real ones. All the examples show how funLOCI works and how it can detect important portions of the continuous domain leading to a segmentation of the functions that classical clustering techniques lack.

\subsection{Case Study 1 - Simulated Data}
\label{simulated_study}
 Simulated data were created using the model proposed by \cite{cremona2018probabilistic}. It generates smooth curves embedding functional motifs (i.e., typical ``shapes'' that may be repeated multiple times within each curve) thanks to the flexibility of B-splines. Considering a B-spline basis $\{\phi_l\}^{L}_{l=1}$ of order $n$, with equally spaced knots $t_1, \dots, t_{L-n+2}$, the model defines each curve as $x(t) = \sum_{l=1}^{L}c_l\phi_{l}(t)$ where $c_{l} \in \mathbb{R}$, $l=1,\dots, L$. Thanks to the fact that $n$ controls smoothness and complexity of $x$, and each basis function has compact support, it is possible to define a functional motif of length $T$ fixing the values of $n$ coefficients $c_{m,i}, \dots, c_{m,i+n-1}$ and repeating them multiple times within the same curve or across different curves. Instead, the background coefficients are randomly sampled out from a $Beta$ distribution. 

We generated a set of 20 curves embedding 4 functional types of local clusters, type A, B, C, and D. Coefficients defining the four types are randomly generated from a $Beta(0.45, 0.45)$ distribution rescaled to $[-15, 15]$. The four types of loci have different lengths and different shapes. Each curve can embed each local cluster zero, one or two times. 

\begin{figure}
\centering
\includegraphics[width=0.8\textwidth]{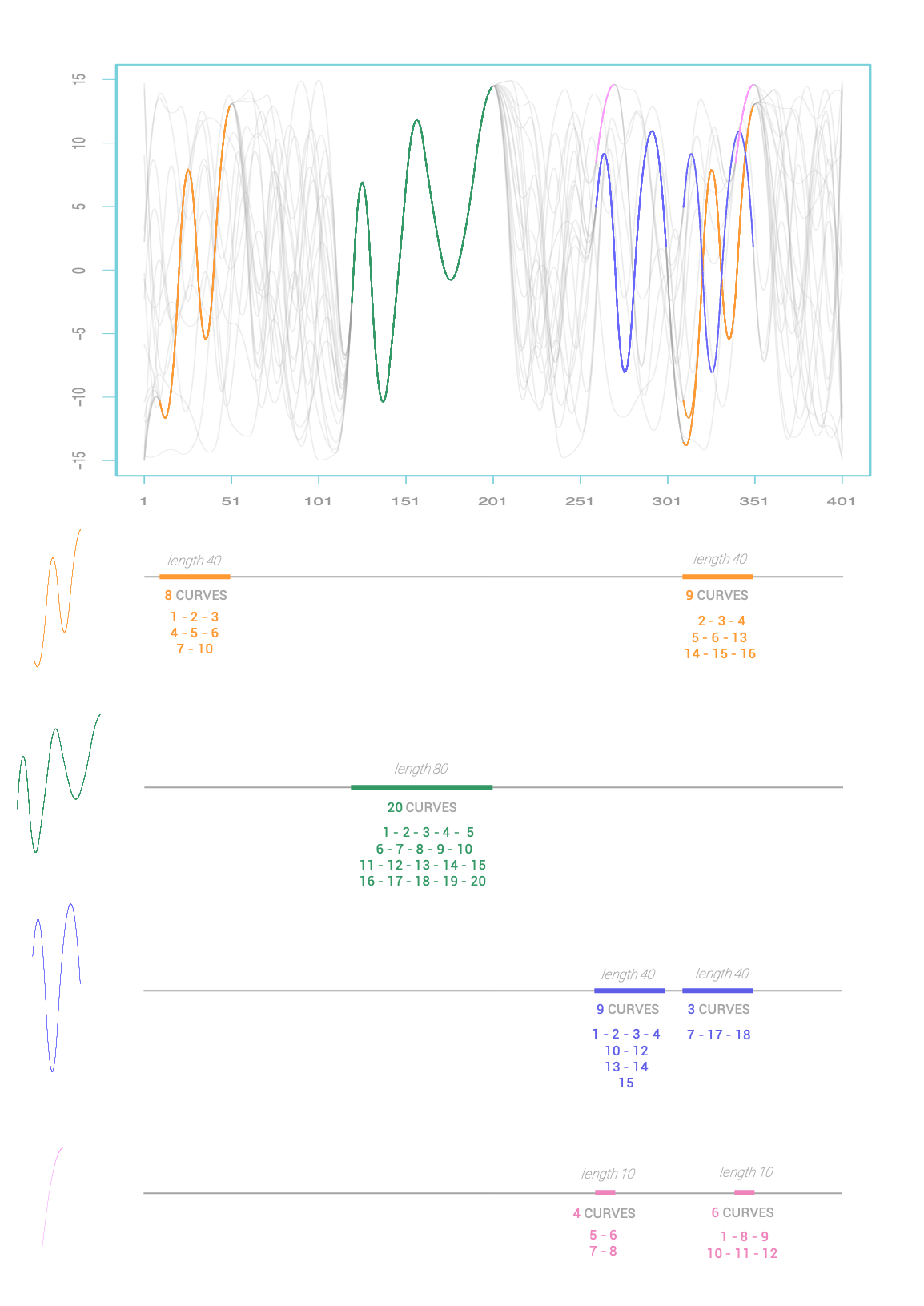}
\caption{The simulated data with $\sigma = 0$. Each local cluster is highlighted with a different color and described in its details: position, length (number of points), number and ids of the curves embedding the locum.}
 \label{fig:motifs}
\end{figure}

\begin{figure}
\centering
\includegraphics[width=0.8\textwidth]{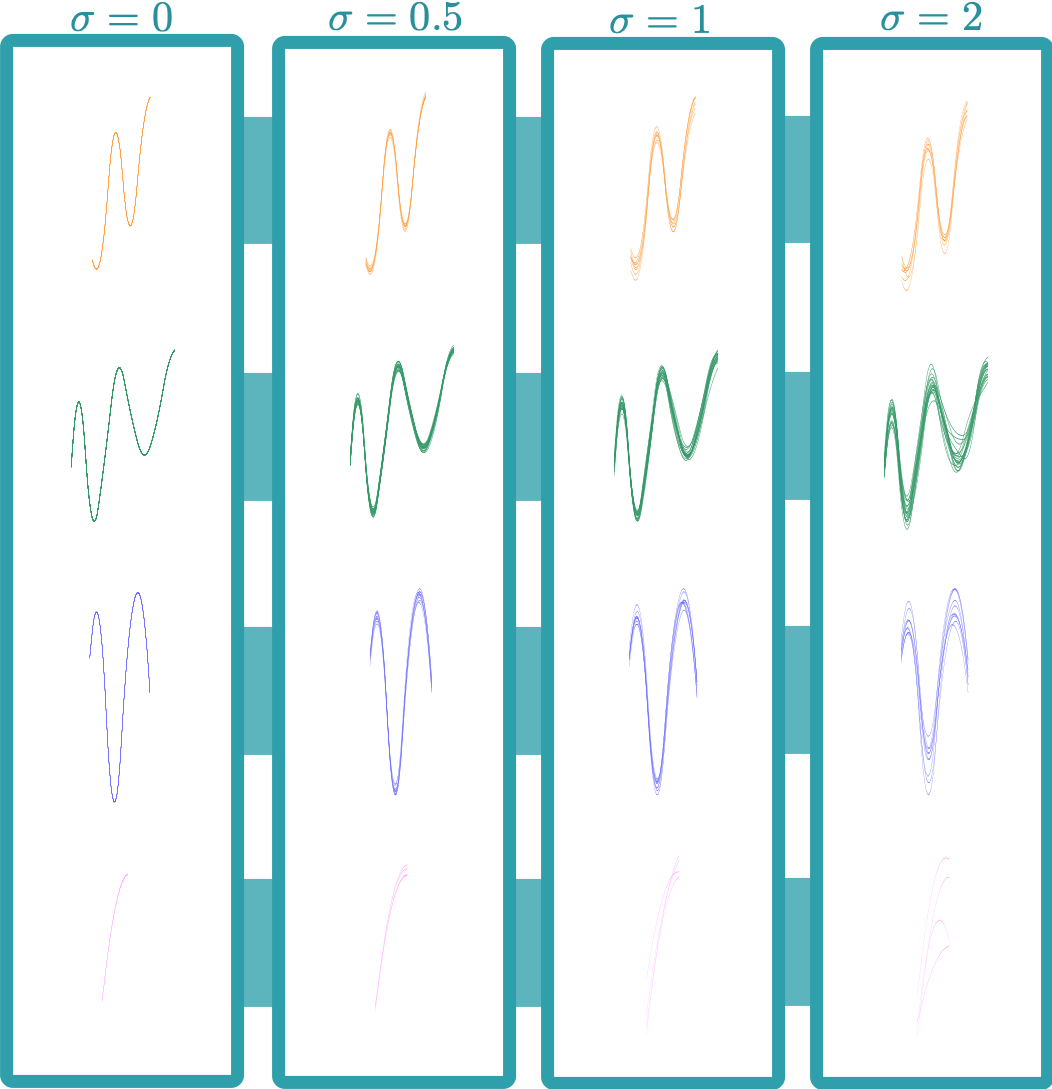}
\caption{The four types of local clusters with different levels of noise $\sigma$. Each row presents a different type: type A is in the first row, Type B in the second one, type C in the third, and type D in the fourth. Each column presents a different $\sigma$: the first column has $\sigma=0$, the second one $\sigma=0.5$, the third one $\sigma=1$, and the last one $\sigma=2$. }
 \label{fig:motifs_sigma}
\end{figure}

All the curves have length $l = 400$ and we generate 4 versions of the same data set. Each version presents the local clusters in the same location but with a different level of noise $\sigma = (0, 0.5, 1, 2)$. 
The $20$ curves for $\sigma = 0$ are shown in Figure \ref{fig:motifs}, with details about the local clusters. Type A and C are defined on a sub-interval of length 40, type B is the longest locum with 80 points, while type D is the local cluster with the shortest interval (10 points). The effect of the level of noise $\sigma$ on the local clusters is shown in Figure \ref{fig:motifs_sigma}.

\begin{figure}
\centering
\includegraphics[width=0.6\textwidth]{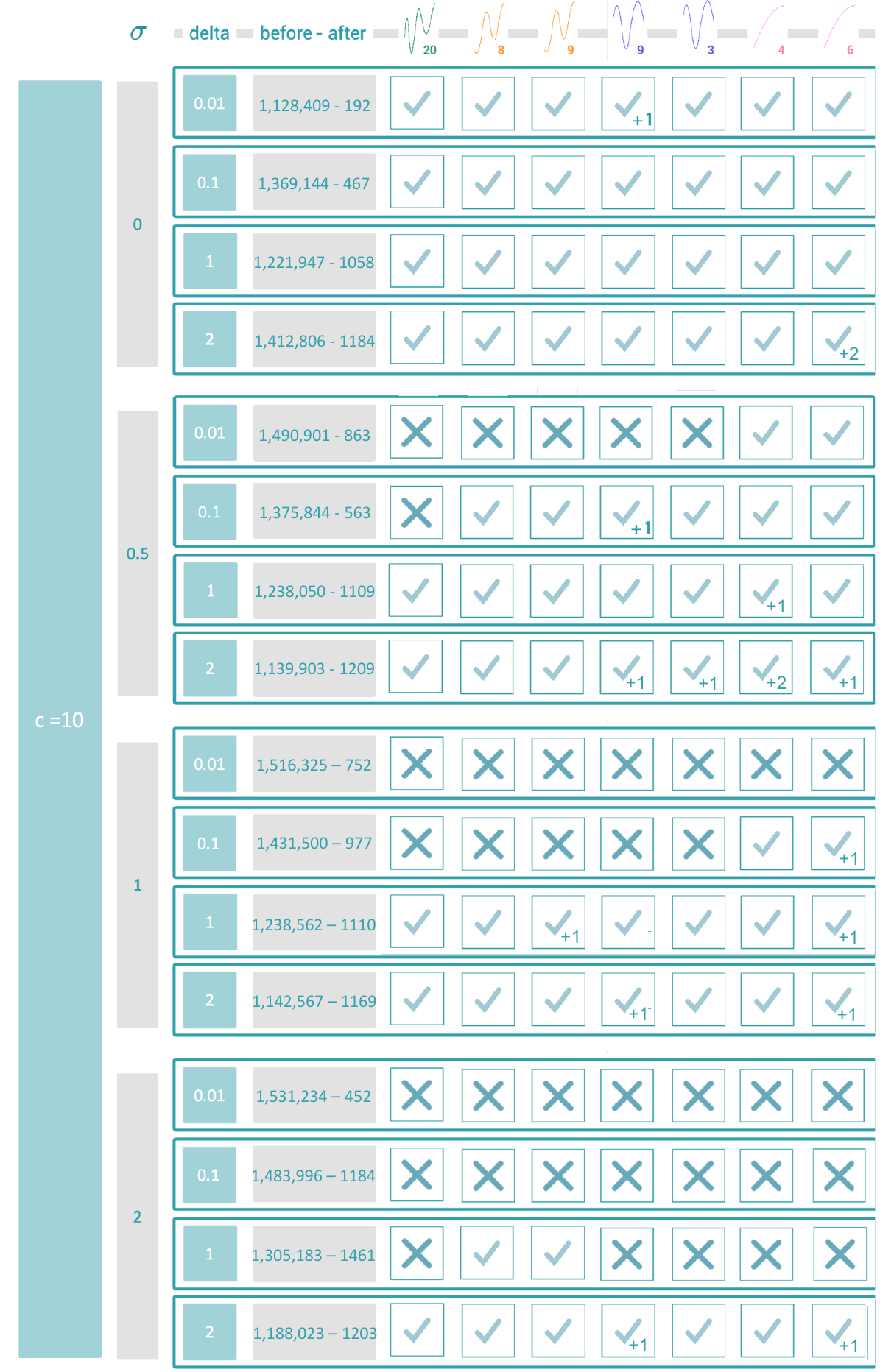}
\caption{For $c=10$ and every different value of $\delta$ and $\sigma$, we listed the number of functional loci before and after the \textit{Tasting}, and for every embedded local cluster if it was correctly detected or not. If it was found with more curves than the original ones we simulated, the number of the extra curves is displayed. Otherwise, if it was found with some missing curves, the number of the missing curves is shown.}
 \label{fig:c10}
\end{figure}

The main aim of this simulation is to detect the four types of loci at different levels of $\sigma$. 
We run the algorithm 8 times: for every data set, we try $c = {10,20}$ in the \textit{Lotting} and $\delta = {0.01, 0.1, 1, 2}$ in the \textit{Harvesting}.
Then, we perform the \textit{Tasting} to reduce the number of local clusters to the minimum by deleting shifted and overlapped results. In this way, we passed from 1-1.5 million of possible candidates to 100-1500 (a reduction of 90-99$\%$). 
Figure \ref{fig:c10} summarizes the results for $c=10$ in different scenarios.

\begin{figure}
\centering
\includegraphics[width=0.4\textwidth]{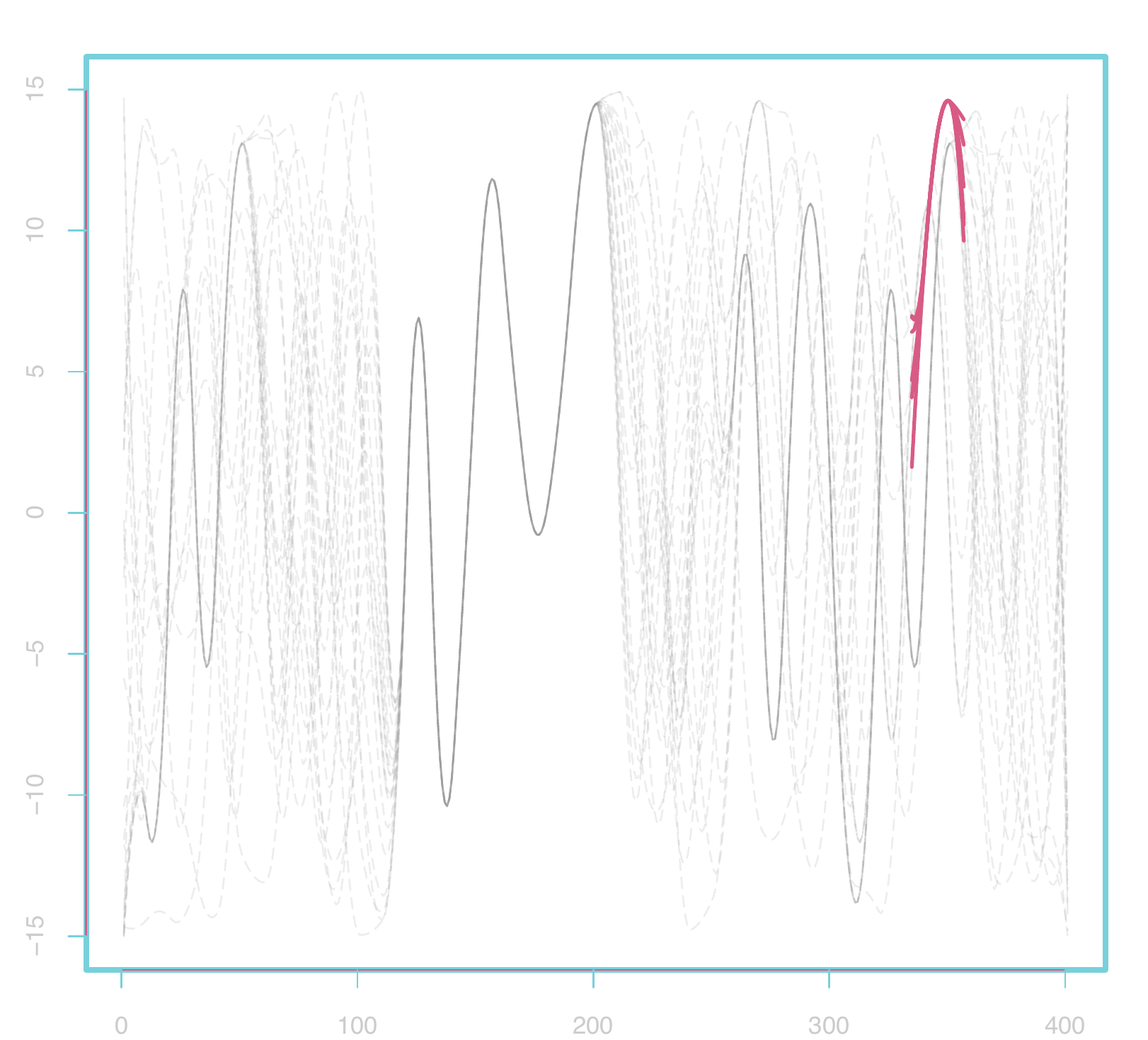}
\caption{When $c=20$ and $\delta=2$, the Type D local cluster composed of 6 curves and $\mid S_w \mid =10$ with $\sigma=0$ is identified by a local cluster with $H-score=0.395$ and $\mid S_{w} \mid > 20$.}
 \label{fig:extended_motif4}
\end{figure}

Fixing $\sigma$, we can notice that the number of loci before the \textit{Tasting} decreases as $\delta$ increases. On the contrary, the number of loci after the \textit{Tasting} increases with some exception (e.g., $\sigma=0.5$ and $\delta=0.1$, $\sigma=2$ and $\delta=1$).
Instead, the choice of $c$ regulates the minimum length of the sub-interval $S_{w}$ and, for instance, when $c=20$, the algorithm can detect only local clusters having $\mid S_{w} \mid \geq 20$. For this reason, funLOCI struggles at finding local clusters having $\mid S_{w} \mid \leq 20$, like the Type D one. Precisely, the algorithm succeeds only if the local cluster is embedded in a larger one with $\mid S_{w} \mid \geq 20$. It usually happens for higher values of $\delta$ ($\delta={1,2}$) and lower values of $\sigma$ ($\sigma = {0,0.5,1}$). Figure \ref{fig:extended_motif4} shows the Type D local clusters detect when $c=20$, $\delta=2$, and $\sigma=0$.

\begin{figure}
\centering
\includegraphics[width=0.8\linewidth]{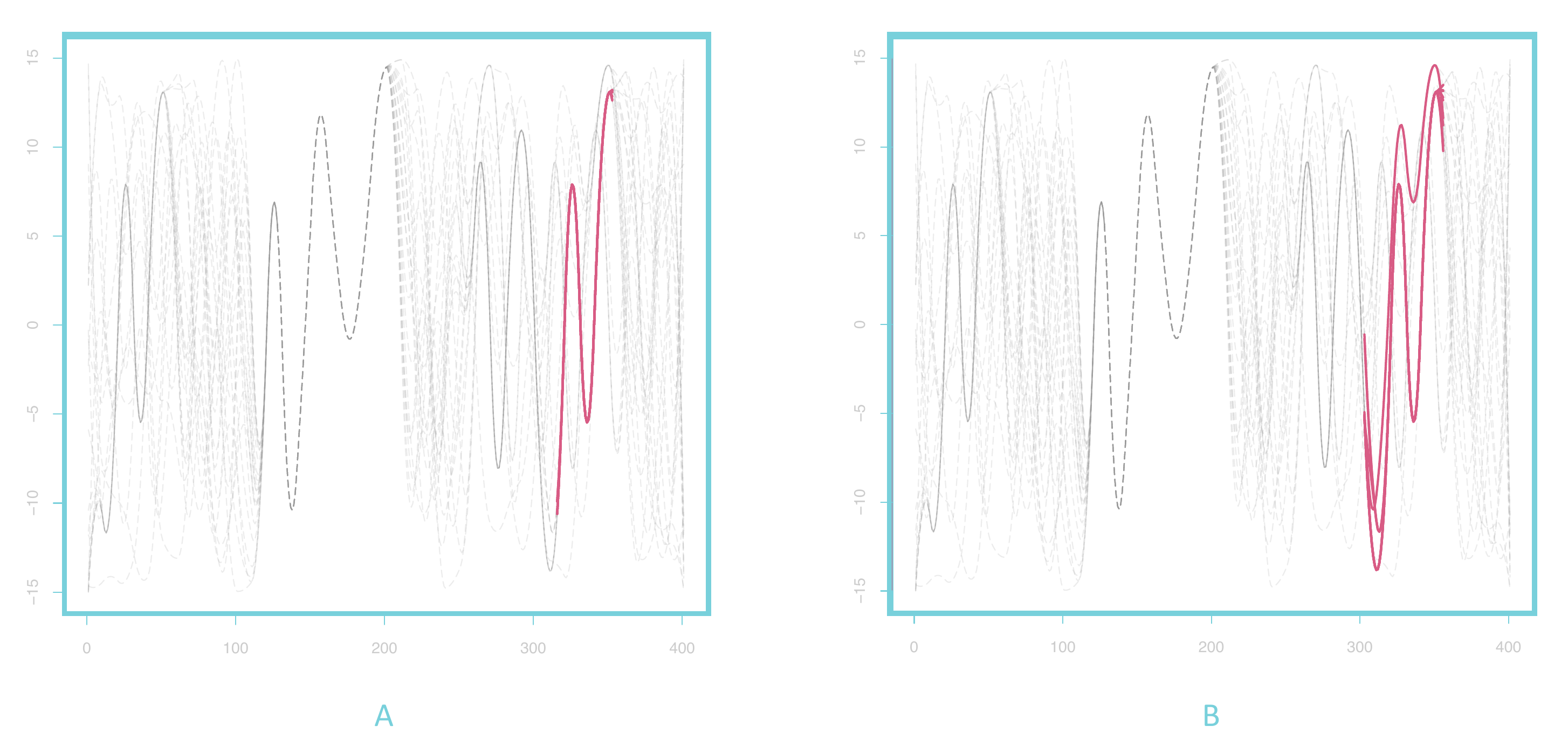}
  \caption{The $\sigma=0$ Type B local cluster, found by funLOCI with $c=10$ when: (A) $\delta=0$; (B) $\delta=2$; Using a larger $\delta$-threshold the local cluster is defined on a larger sub-interval and it counts one extra curve.}
  \label{fig:comparison_sigma}
\end{figure}

The $\delta$ threshold identifies the maximum H-score of the resulting local clusters, affecting the adherence to the unknown additive models. In general, lower values of $\delta$ lead to more cohesive results. However, higher values of $\delta$ could still lead to the detection of the local cluster but usually with more curves and a longer sub-interval (see Figure \ref{fig:comparison_sigma}).
Instead, when $\delta < \sigma$, the algorithm usually fails and it detects only portions of the local cluster. This is because funLOCI is looking for results with an H-score lower than the one characterizing the locum.
However, generally speaking, using the correct parameters the algorithm usually succeeds.

\begin{figure}
\centering
\includegraphics[width=0.4\textwidth]{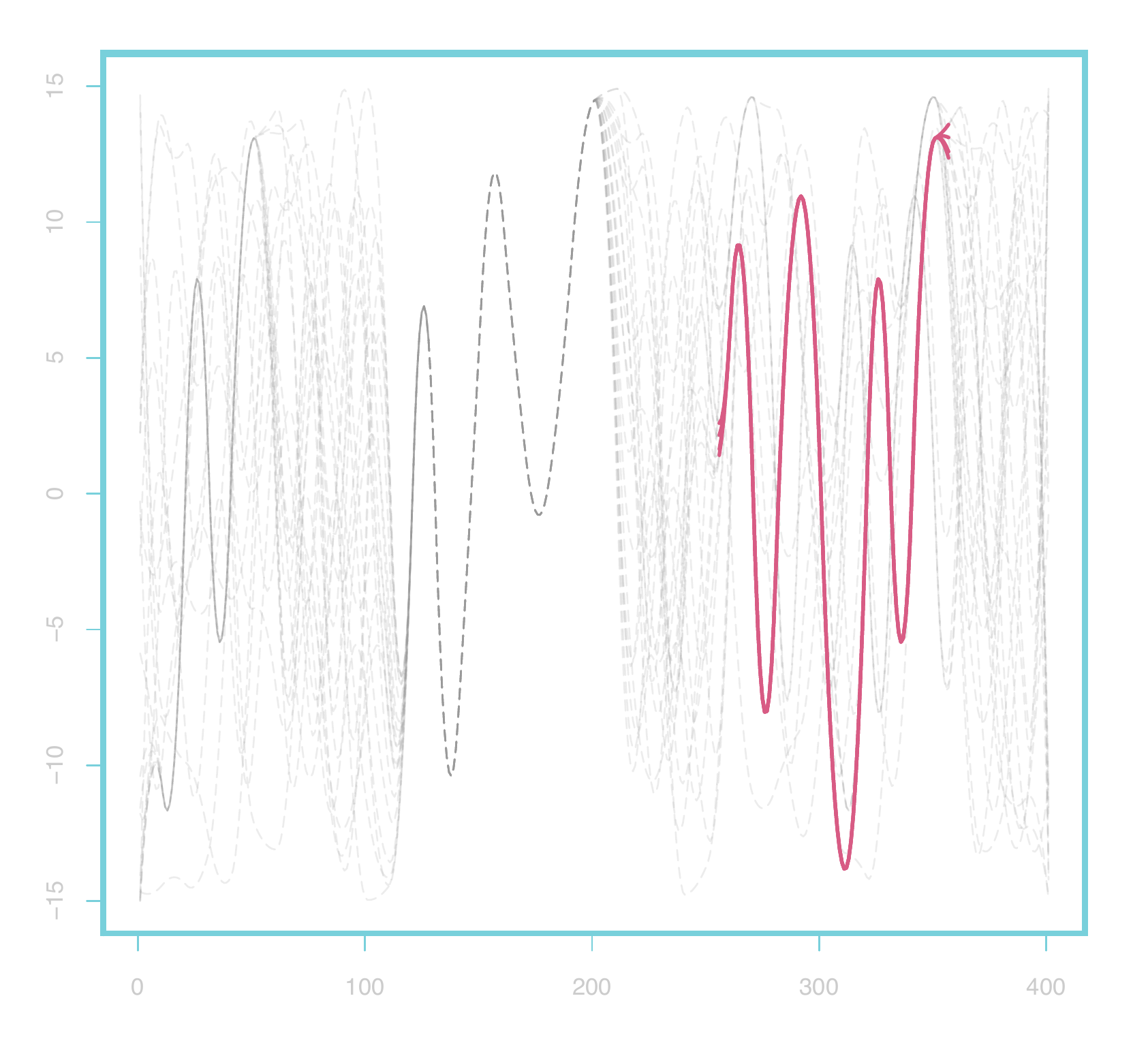}
\caption{A local cluster composed of 4 curves created by the random generation of the background. It is the result of the concatenation of Type D and Type A loci.}
 \label{fig:extra_motif}
\end{figure}

Another interesting aspect to take into account is that the algorithm detects many other functional local clusters than the ones we embedded. This is due to the fact that the random generation of the background of the simulated data can create new loci we were not aware of. An example is the one presented in Figure \ref{fig:extra_motif}, resulting from the linkage of the Type C and Type A local clusters for 4 curves.

 Sometimes, the randomly generated loci can even be more interesting (in terms of number of elements, sub-interval length and H-score) than the ones we embedded (see Type D local clusters). In other cases, one local cluster can be very interesting according to one criterion (e.g., H-score) but not to others. For these reasons, we believe that the use of interactive visualization tools or multiple ordering criteria can be extremely useful in the identification of local clusters.

\subsection{Case Study 2 - Aneurisk Project}

\begin{figure}
\centering
\includegraphics[width=0.48\textwidth]{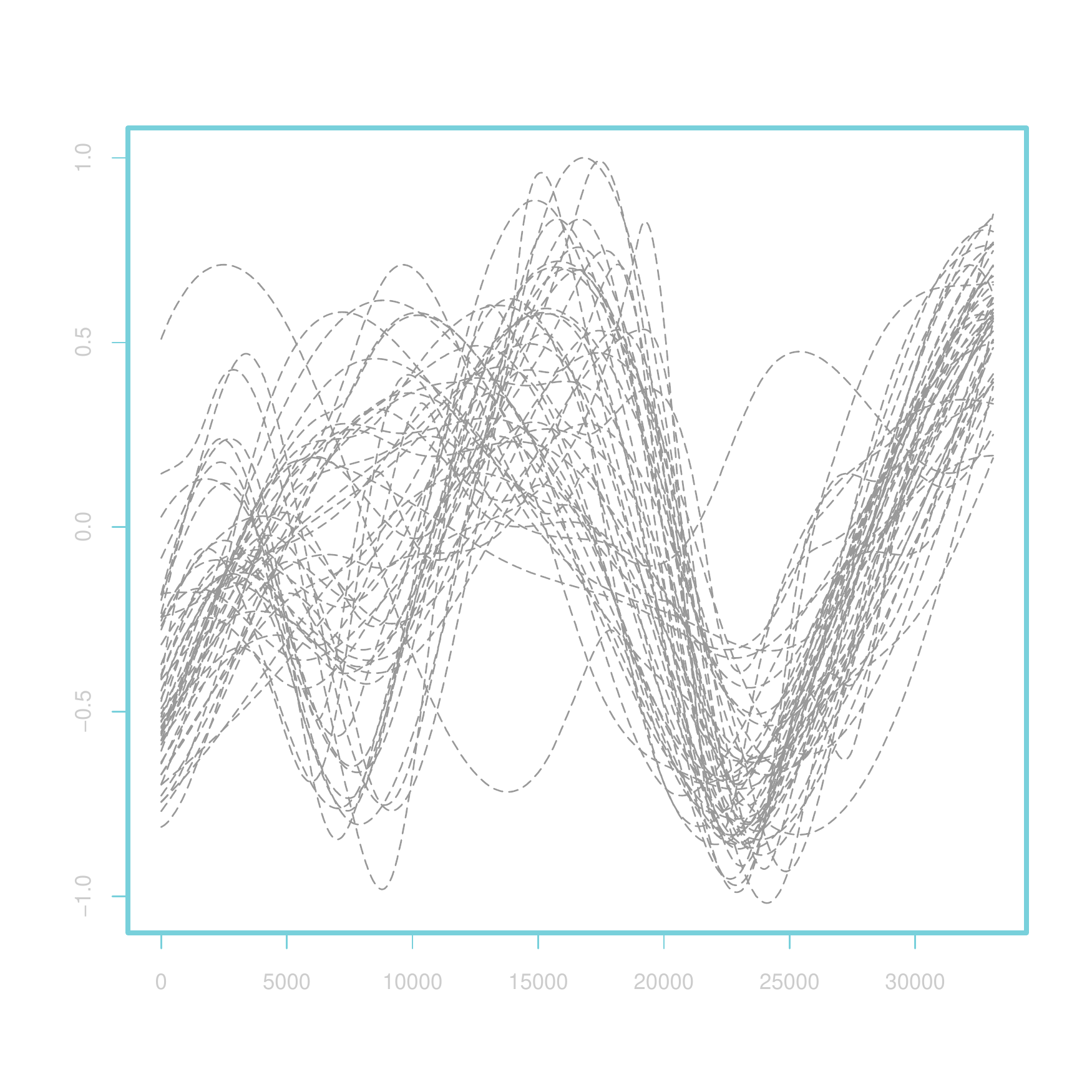}
\caption{The 50 $z$-first derivative curves of the ICA vessel centerlines after registration}
 \label{fig:aneudata}
\end{figure}

The AneuRisk65 data have been collected within the AneuRisk project, a multidisciplinary scientific endeavour aiming at investigating the role of vessel morphology, blood fluid dynamics, and biomechanical properties of the vascular wall, on the pathogenesis of cerebral aneurysms. 65 subjects took part to the project. The data present, for each subject, both raw and preprocessed information about the Inner Carotid Artery (ICA), described in terms of vessel centerline and radius profile. All data are available at \url{https://statistics.mox.polimi.it/aneurisk/}, the official website of the project. In this case study, as in the paper by Passerini et al. (\citeyear{passerini2012integrated}), we consider 50 $z$-first derivative vessel centerline curves obtained after registration. Each curve is defined in the last portion of the ICA and it is 33101 data points long (Figure \ref{fig:aneudata}).

If in the \textit{Lotting} step $c = \mid T \mid$, the algorithm performs mean squared residue score-based DIANA to the whole domain. 

\begin{figure}
\centering
\includegraphics[width=0.7\textwidth]{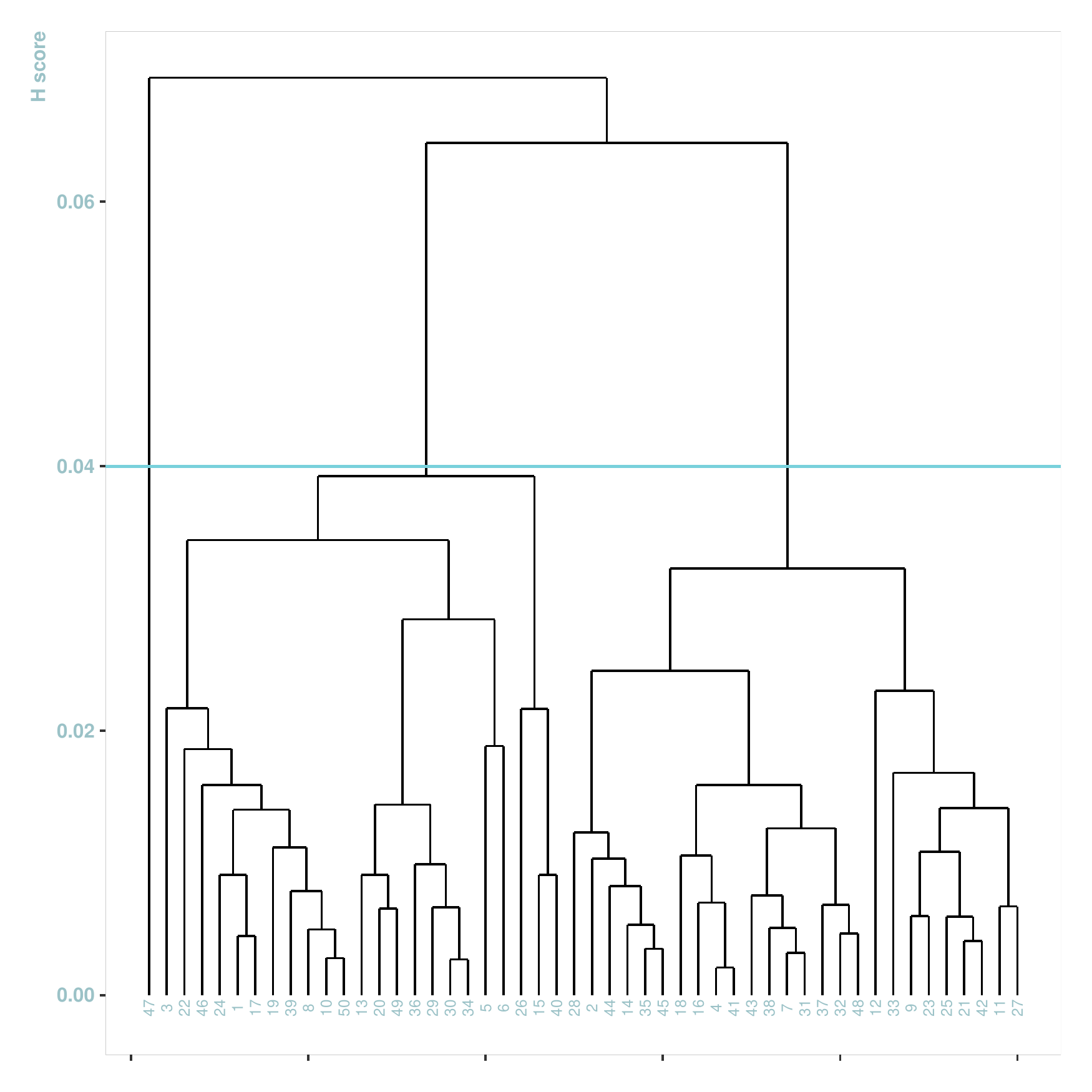}
\caption{The dendrogram obtained by the mean squared residue score-based DIANA on the the whole domain.  $\delta=0.04$ cuts the dendrogram in $3$ groups.}
 \label{fig:aneudendro}
\end{figure}

As one can see from the dendrogram in Figure \ref{fig:aneudendro}, setting for instance $\delta = 0.04$, $3$ global clusters are identified.

\begin{figure}
\centering
\includegraphics[width=0.8\textwidth]{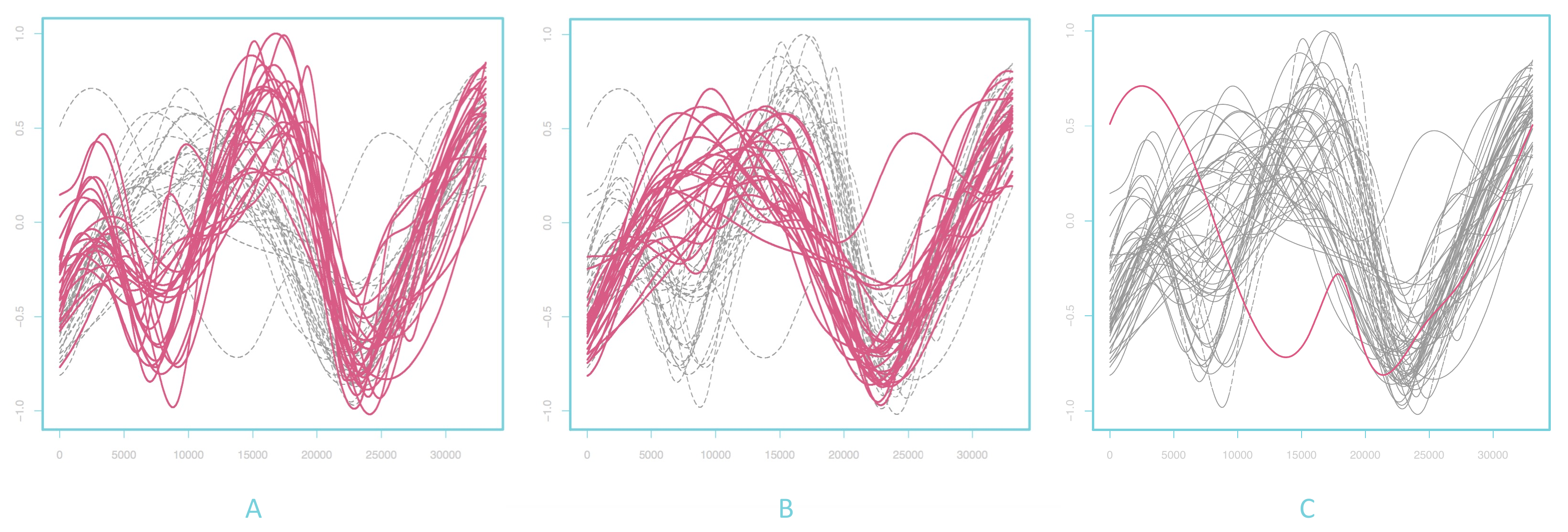}
\caption{\textbf{A:} $\Omega$ shape vessels; \textbf{B:} $S$ shape vessels; \textbf{C:} Singleton composed by an outlier resulting from a registration problem in the original data.}
 \label{fig:omega_s_outlier}
\end{figure}

These results identify the typical $\Omega$ and $S$ shape groups of the vessel centerline of the inner carotid artery and a cluster composed by a single outlier (Figure \ref{fig:omega_s_outlier}). These results are coherent with the medical literature (\cite{huber1982krayenbuhl}) and previous works (\cite{sangalli2009case} and \cite{sangalli2012joint}).

However, the main scope of funLOCI is to find functional local clusters to discover dynamics that global clustering is not able to highlight. Being satisfied by the 3 clusters aforementioned, we set the same $\delta$-threshold ($\delta = 0.04$) for the identification of local clusters. Precisely, we create 4357 sub-intervals, i.e. all the possible ones identified on a grid of step $250$ ($\bar{a} = \{1, 251, \dots \}$ with minimum length $c = 500k, \hspace{3px} k = 1,2,\dots$, ($\bar{c} = \{500, 1000, \dots, 33101\}$). funLOCI was able to detect $8007$ local clusters. The \textit{Tasting} procedure reduces that number to $67$, performing a reduction of the $99.2\%$.

\begin{figure}
\centering
\includegraphics[width=0.55\textwidth]{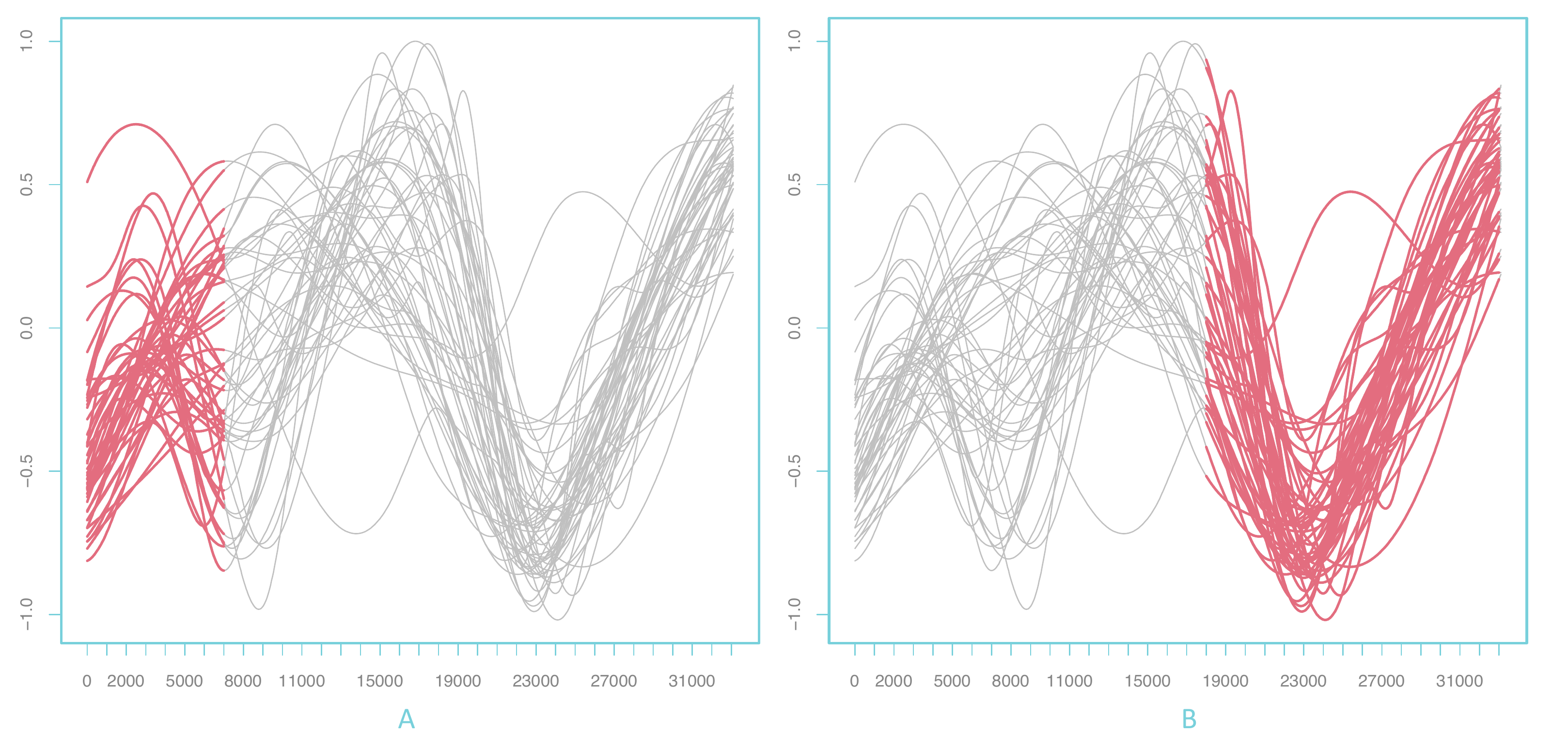}
\caption{All the $z$-first derivative curves of the ICA vessel centerlines compose one single local cluster both at the beginning (panel A) and at the end (panel B) of the domain.}
 \label{fig:aneu_alltogether}
\end{figure}

Three local clusters which show the potential of the method are here presented. Figure \ref{fig:aneu_alltogether}.A shows that in the sub-interval $(1,7001)$,approximately the first portion of the domain $T$, all the curves belong to the same group. In the panel B, we can see that the same curves are grouped together also in the interval $(7001,33101)$. Therefore, the identification of the $S$ and $\Omega$ shape groups is not to be accounted for the first or the second half portion of the vessels.

\begin{figure}
\centering
\includegraphics[width=0.55\textwidth]{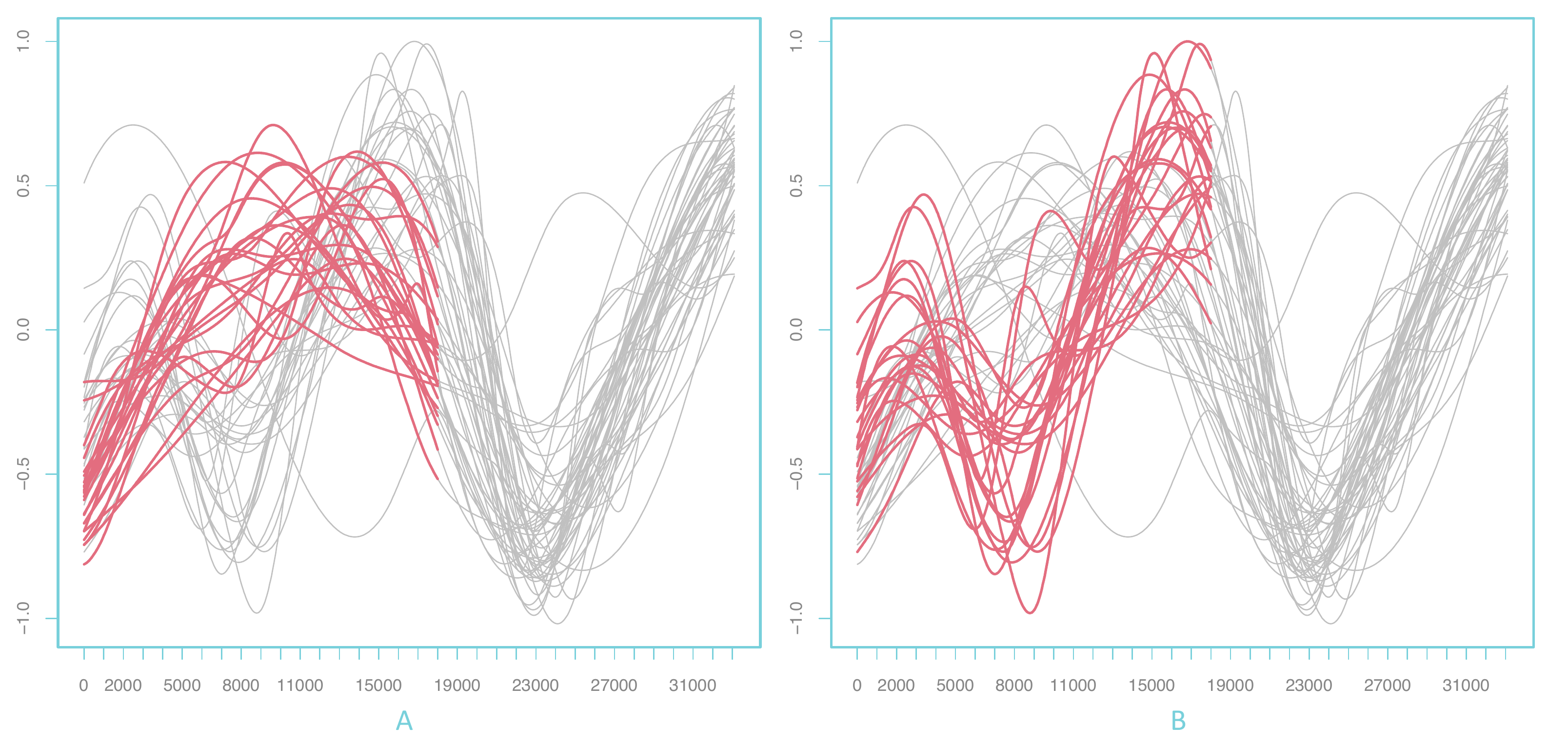}
\caption{All the $z$-first derivative curves of the ICA vessel centerlines compose one single local cluster both at the beginning (panel A) and at the end (panel B) of the domain.}
 \label{fig:from1to18000}
\end{figure}

Let us focus instead on the first half of the domain $T$, precisely on the the sub-interval $S_{w} = \{1, 18001\}$ (see Figure \ref{fig:from1to18000}). There, we identify two local clusters: the first counts all the curves belonging to the $S$-shaped group with the exception of the curve 28, while the second locum presents the curves inside the $\Omega$-shaped cluster plus the curve 28. It means that the partition obtained with the global clustering should be mainly imputed to the first half of the domain $T$. The curve 28 is the only curve that is not correctly assigned to its shape cluster.

A similar situation characterizes the two local clusters identified in the sub-interval $S_{w} = \{7001, 18000\}$, but with three, and not one, misassigned curves (curves 14, 28, 45).

\section{Discussion}

In this paper, we introduce the functional local clusters identifier algorithm (funLOCI). This method allows to identify functional local clusters or functional loci, defined as subsets of curves characterized by similar behaviour (e.g., shape) across the same continuous subset of the domain.  Taking inspiration from the most comprehensive formulation of multivariate bicluster, the definition of these objects is based on an additive model taking into account the shapes of the curves. To evaluate the goodness of a functional locum, one has to evaluate the adherence to the corresponding additive model which is, however, unknown a priori. The validation score that we use is the functional version of the mean squared residue score or H-score and it is one of the core element of the funLOCI algorithm. 

funLOCI is three-step algorithm that uses a brute-force approach and it is characterized by an high flexibility. All the three steps are, indeed, easily generalizable: the structure of funLOCI could be modified not only in the \textit{Flowering} step, where different hierarchical clustering algorithms can be used, but also in the \textit{Harvesting}, where one can employ different collection strategies. All the R scripts and libraries running the method can be downloaded in the GitHub repository at \url{https://github.com/JacopoDior/funloci}.

Nonetheless, there are still open points.
In the \textit{Harvesting}, tuning the parameter $\delta$ can be problematic, especially when there is lack of guidance from the domain expert or when the H-score of all the curves defined in a particular sub-interval largely changes from one sub-interval $S_{w}$ to another. As an alternative, we propose the implementation of $\delta^{\%}$ or the use of elbow method. Other alternative strategy can be implemented in the next future.

As already observed, it could also be really difficult to identify the most interesting local clusters when the amount of candidates is overwhelming. For this reason, the \textit{Tasting} procedure is introduced. This step reduces the results to the minimum by deleting nested and shifted overlapping loci. Another way to further improve the identification of the best results is to order the local clusters according to interesting criteria such as H-score or length of the sub-interval, or exploring the results using interactive visualization. For this reason, we proposed an interactive visualization tool which permits to the user to manually explore the results. New features for the tool are currently under development.

In addition, funLOCI can only discover subsets of curves with similar behaviour across the same continuous subsets of the domain. Consequently, it can not find similarity across different sub-intervals. Therefore, the use of an alignment strategy could be a possible direction of generalization in order to detect similar functions across different sub-intervals.
\\
\\
\textbf{Code Availability:} The R code implementing the procedure is available at \url{https://github.com/JacopoDior/funloci}.
\\
\\
\textbf{Conflict of interest:} The authors declare no competing interests.
\printbibliography

\end{document}